\documentclass{micro-econ-thesis}
\usepackage{graphicx}
\graphicspath{{figures/}}
\usepackage{setspace}

\usepackage[giveninits=true, uniquename=init, hyperref=true, backend=biber, style=ieee, sorting=none]{biblatex}
\usepackage{pdfpages}
\usepackage{stackengine}
\usepackage{amssymb}
\usepackage{booktabs}
\usepackage{multirow}
\usepackage{rotating}
\usepackage{amsmath}
\usepackage{commath}
\usepackage{IEEEtrantools}

\newcommand{\argminD}{\arg\!\min}

\usepackage[utf8]{inputenc} 
\cleardoublepage
\begin{document}

\title{A Tutorial on Adversarial Learning Attacks and Countermeasures }
\author{Cato Pauling, Michael Gimson, Muhammed Qaid, Ahmad Kida and
\and  Basel Halak
\and School of Electronics and Computer Science, University of Southampton, UK
\and email: basel.halak@soton.ac.uk}
 
\maketitle


\abstract{
Machine learning algorithms are used to construct a mathematical model for a system based on training data. Such a model is capable of making highly accurate predictions without being explicitly programmed to do so. These techniques have a great many applications in all areas of the modern digital economy and artificial intelligence. More importantly, these methods are essential for a rapidly increasing number of safety-critical applications such as autonomous vehicles and intelligent defense systems. However, emerging adversarial learning attacks pose a serious security threat that greatly undermines further such systems. The latter are classified into four types, evasion (manipulating data to avoid detection), poisoning (injection malicious training samples to disrupt retraining), model stealing (extraction), and inference (leveraging over-generalization on training data). Understanding this type of attacks is a crucial first step for the development of effective countermeasures. The paper provides a detailed tutorial on the principles of adversarial machining learning, explains the different attack scenarios, and gives an in-depth insight into the state-of-art defense mechanisms against this rising threat
.}


 

\onehalfspacing 

\section{Introduction}
\label{sec:background}

Our world is becoming increasingly dependent on machine learning for making automated decisions. With rapid rise in malicious cyberattacks, where attackers’ incentives are to extract and manipulate the results and models generated by machine learning algorithms, the need for efficient and practical defences for these systems is crucial \cite{regressionattack}.

Intelligent Security Detection Systems (ISDS) are systems designed to identify and mitigate malicious activity. Anomaly detection is a common example that uses machine learning (ML) algorithms to derive a model of trustworthy behaviour. The current activity is then compared to the normal behaviour learnt by the model, which is then able to classify the activity as normal or anomalous, triggering the appropriate response. These security systems themselves may be vulnerable to adversarial learning attacks which target the ML algorithms they are based upon.

Adversarial learning attacks against machine learning systems exist in an extensive number of variations and categories; however, they can be broadly classified: attacks aiming to poison training data, evasion attacks to make the ML algorithm misclassify an input, and confidentiality violations via the analysis of trained ML models. Many attacks can have their application adapted to both deep learning systems and more traditional machine learning models. Poisoning attacks have a prerequisite of access to the training dataset of the model, before it is trained and have attack implementations for classical machine learning methods including linear regression \cite{regressionattack} and Support Vector Machines \cite{svmattack}. 

The remainder of this paper is structured as follows. Section 2 discusses the threat modelling approaches for adversarial machine learning. Attack mechanisms are discussed in section 3, followed by a review of countermeasures in section 4. Conclusions are drawn in section 5.

\section{Threat Modelling for Adversarial Leaning Attacks}
\label{sec:general principles}

\subsection{Adversarial Threat Model}
\label{subsec:generalthreatmodel}

Papernot et al. \cite{papernot2016science} provided a summary of the type of adversarial approaches machine learning algorithms are vulnerable to and offered a comprehensive ‘unifying threat model’ for these algorithms and the encompassing systems that use them. The life cycle of a machine learning system is treated as a generalised data processing pipeline. The entire pipeline as a whole is considered, with a specific focus on the two distinct phases of a machine learning model: the training phase and the inference phase. They divide the threat model into three key components: the attack surface, adversarial capability and adversarial goals.

The attack surface of a ML system describes the point in which the attack takes place. This can be described with respect to the pipeline. For example, such a system may begin by first collecting the data it wishes to feed into the model, from sensors or data repositories. This data usually will need to be processed before being fed into the model, which produces an appropriate output. The model output is then communicated to an encompassing system in order to be acted upon. In such a system, the adversary may want to interfere with collection or processing of input data, corrupt the model itself or tamper with the model’s output. Each of these are an appropriate attack surface. Most adversarial learning attacks will occur during the input data collection and processing in order to exploit the vulnerabilities in the model without corrupting it. This can be done by providing specific information for the sensors to pick up, or adding perturbations to images after being collected by the sensors.

The capability of the adversary also plays a large role in defining the threat model. The definition of a system’s vulnerabilities is also made with regard to the strength of the attacker, who may have greater access to the target system and its data, and thus a greater ability to inflict damage to the system. The capabilities of an adversary outline the sort of attacks available to them. These relate to what part of the process they have access to, be it the training phase or the interference phase.

If the adversary has the ability to insert themselves into the training phase, they have the capability to learn or influence the training of, and therefore corrupt, the model itself. The simplest form of attacks in this phase involves having access to the training dataset, in full or partially. With this, provided the quality and quantity of data is sufficient, the adversary can train a substitute model on the same dataset in order to test adversarial inputs, and validate their success. Broadly however, there are two general strategies available in the training phase. The first of these is to meddle with the training set. An adversary can insert malicious samples into the training set to sabotage what the model learns and throws the model off. This form of altering the training set it referred to as \textit{injection}. Similarly, \textit{modification}, involves directly altering the training set. Alternatively, the adversary may be able to tamper with the actual learning logic of the model. These attacks known as logic corruption, are extremely powerful and would be very tough to defend against. 

In the inference phase however, attacks do not attempt to corrupt the model itself, but rather to fool it into producing incorrect outputs, by exploiting inherent vulnerabilities. Alternatively, they may try to perform reconnaissance, merely gathering evidence of the characteristics. Inference time attacks, generally, can be split into white-box or black-box attacks depending on the adversary’s access to information on the target model’s architecture and therefore their ability to exploit it. 

The final component which needs to be considered in Papernot’s threat model \cite{papernot2016science} are the adversarial goals. They model the goals of potential adversaries using the classic cyber security mode, CIA, which defines the three characteristics a system aims to uphold, and an adversary may want to target: confidentiality, integrity and availability. Attacks on the confidentiality of the system correspond to attempting to extract information on the model itself or its training data. For example, in the context of a financial system, the model itself may be regarded as confidential intellectual property, or likewise in the context of a medical system, the training set may contain confidential data used to train it. Attacks on integrity aim to cause the model to behave in an unintelligent and incorrect manner thereby undermining the integrity of the model. For example, causing the misclassification of certain inputs to the model or reducing the confidence the model has in its output. Integrity is crucial to machine learning, and therefore integrity metrics, like model accuracy are often the primary focus of performance metrics. In the same way, attacks on availability look to thwart access to output of such models. One way this can be achieved is by targeting the model’s quality and consistency as a model which is erratic and unreliable may consequently become useless. Attacks in this domain will use similar methods as those used to damage the integrity of the model. Attacks on availability also include attempts to reduce the access to the model, via denial-of-service attacks, or model’s performance metrics such as speed so it becomes inconvenient to use. 

\subsection{Intelligent Security Detection System Threat Model}
\label{subsec: ISDSthreatmodel}

Intelligent Security Detection Systems which employ similar machine learning models fit into this general threat model are therefore vulnerable to adversarial learning attacks. A common approach for such systems is to learn a model of trustworthy behaviour, from which current behaviours can be compared to. Therefore, these systems can generally be attacked in the training stage by poisoning the training set or by logic corruption. Similarly, attacks in the inference stage would aim to evade detection or potentially trigger false positives. These false positives may trigger unnecessary reactions by the system, potentially causing unwanted responses such as the wiping of important data. A stream of false positives may also strain the system, reducing its performance, and as such, its availability.

\subsection{Attack Attributes}
\label{subsec:attcksthreatmodel}

Following on from the general threat model described earlier, the adversarial learning threat model can be broken down into a number of more granular attributes. These attributes are adversarial falsification, adversarial knowledge, adversarial specificity, attack frequency and are selected appropriately depending on the scenario of a specific attack and its accompanying assumptions and quality requirements \cite{8611298}. 
\subsubsection{Adversarial falsification}
\label{subsubsec:Adversarial falsification}

\noindent Adversarial falsification distinguishes between whether the adversary aims to produce a false positive attack or false negative and what this means for the target system. A false positive attack results in negative samples being incorrectly classified as positive. In the case of an intrusion detection system this may result in the inappropriate triggering of a response to the perceived attack such as deletion of non-threatening or important information. Similarly, the existence of a stream of false positives which the system has to deal with may negatively impact the performance of the model. On the other hand, false negative attacks, also known as evasion attacks, work in the reverse, using the same example of an intrusion detection system, the true intruder or malware is classified as benign and left completely undetected. 

\subsubsection{Adversarial specificity}  \hspace{0pt} \par
\label{subsec:specificity}
\noindent Adversarial specificity differentiates between targeted and non-targeted attacks and usually relates to the case of a multiclass classification. Targeted attacks look to guide the output of the model in a certain direction to a specific class. For example, duping the model into predicting all adversarial examples as one class. However, this isn’t usually necessary and non-targeted attacks merely aim to classify an adversarial example as any class other than the original. These are therefore easier to implement and are either realised by reducing the probability the model classifies correctly or by selecting one from numerous targeted approaches, the one with the slightest perturbation. 

\subsubsection{Adversarial frequency}  \hspace{0pt} \par
\label{subsec:frequency}
\noindent Attack frequency concerns whether the attack is a one-time approached or an iterative one, which updates a number of times to better optimise the attack. The latter tend to perform better, however come with extra cost, requiring greater computational time. In some cases, the one-time attacks are sufficient or even the only feasible option. 

\subsubsection{Adversarial Knowledge: White-box vs Black-box}  \hspace{0pt} \par
\label{subsec:whiteboxblackbox}

\noindent Adversarial knowledge relates to adversarial capabilities and more particularly refers to the case of inference time attacks. Attacks in this case can be broadly split two distinct categories based on the level of detail they require of the target model, to be performed. A white box attack is one which requires an in-depth understanding of the target model. In this way, the mechanics of the model are transparent. The attacker has full knowledge of the system and has access to vital information including details such as the network architecture, parameters, hyper parameters, training data as well as the ability to gather gradients, and prediction results \cite{8611298}. On the other hand, a black box attack attacks the model with no knowledge of the inner workings of the model. A black box attacker simply requires access to the model in order to query it to produce the prediction result \cite{papernot2017practical}. \\

In the context of a black box attack, an attacker may be able to produce a surrogate model which acts in accordance with the inputs and outputs of the target model. In this way, the attacker has full access to the inner workings of their surrogate model and can therefore successfully conduct white box attacks on it by creating adversarial samples using this knowledge. These adversarial examples can then be used to conduct a black box attack against the target model. \\

More recently, a third category has emerged as an alternative to both white-box and black-box approaches. Grey-box attacks assume knowledge of the target model’s basic architecture without detailed information such as the network weights \cite{Vivek_2018_ECCV}. An example would be where the target model architecture and training process are known, and a substitute model can be trained to create adversarial examples \cite{REN2020346}. 

\section{Types of Attack}
\label{sec:attacktypes}

Adversarial learning attacks come in different forms. In this section, we review different types of adversarial learning attacks against deep learning architectures. We focus the attacks against deep learning architectures as these architectures are quickly gaining popularity in industrial application. This is because of the exponentially increasing amount of data that is becoming available which is used to train these models. The performance of deep learning models surpasses that of classical machine learning models when trained on large amounts of data \cite{janiesch2021machine}.

\subsection{Fast Gradient Sign Method}
\label{subsec:fgsm}

The Fast Gradient Sign Method (FGSM) was introduced by \textcite{goodfellow2015explaining} as a part of an explanation of the nature of adversarial examples and what makes neural networks so vulnerable to them. This behaviour had previously been demonstrated by \cite{szegedy2014intriguing} whose L-BFGS method exhibited fundamental weaknesses in neural networks and their training algorithms. 
\cite{goodfellow2015explaining} argues that neural networks are vulnerable to linear adversarial perturbation attacks due to their linear nature and that the analytical exploiting of this behaviour allows for a cheap and effective method of perturbation generation. This can be demonstrated by taking the dot product of a weight vector $\mathbf{w}$ and an adversarial example $\mathbf{\bar{x}} = \mathbf{x} + \pmb{\eta}$:

\begin{equation}\label{FGSM1}
    \mathbf{w}^\intercal \mathbf{\bar{x}} = \mathbf{w}^\intercal + \mathbf{x} \mathbf{w}^\intercal \pmb{\eta}
\end{equation}

This perturbation causes the activation to grow by $\mathbf{w}^\intercal \pmb{\eta}$. By assigning $\pmb{\eta}$ to $sign(\mathbf{w})$ we can maximise the effect of the perturbation, subject to the max norm constraint. With this, we can now see that the maximum bound of the change in the activation is

\begin{equation}\label{FGSM2}
    \mathbf{w}^\intercal \pmb{\eta} = \epsilon \cdot \mathbf{w}^\intercal sign(\mathbf{w})
    = \epsilon m n
\end{equation}

where $\mathbf{w}$ has $n$ dimensions and the average magnitude of an element of the weight vector is $m$.

This shows that the activation given by the perturbation grows linearly with the number of dimensions, $n$. Therefore, in adequately high-dimensional space, many infinitesimal changes, each capped at $\epsilon$, can sum up to produce large enough perturbation to the output, able to fool the model. 

FGSM looks to work in the opposite manner to gradient descent in order to maximise the loss. Therefore, we use the cost function used to train the neural network, $J$, to calculate the gradient of the loss with respect to a particular input $x$. This allows us to push each pixel in the direction of the gradient, the direction which most increases the loss. This can be expressed as  

\begin{equation}\label{FGSM3}
    \pmb{\eta} = \epsilon \cdot sign(\Delta_\mathbf{x} \mathbf{J(w,x,y)})
\end{equation}

where $y$ are the targets associated with $x$.

This is then added to the original input to create the adversarial example.

\begin{equation}\label{FGSM4}
    \mathbf{x^*} = \mathbf{x} + \epsilon \cdot sign(\Delta_\mathbf{x} \mathbf{J(w,x,y)})
\end{equation}

This one-step gradient method was then used a basis for a series of similar approaches. The basic iterative method (BIM) (also known as iterative- FGSM) is one such attack \cite{kurakin2017adversarial}. This extends the FGSM by iterating the process, using a reduced step size and clipping pixel values of the transitional results in order to keep them within a defined range of the original input, described as the ‘$\epsilon$- neighbourhood’. This can be expressed as: 

\begin{equation}\label{IFGSM}
    \mathbf{x^*}_{N+1} = \mathnormal{Clip_{X,\epsilon}} \texttt{\{} \mathbf{x}_N + \alpha \cdot sign(\Delta_\mathbf{x} \mathbf{J(w,x,y)}) \texttt{\}}
\end{equation}

In the same work, \cite{kurakin2017adversarial} also explores the idea of pushing a particular misclassification rather than only pushing in the direction of the target class. This can result in more extreme and interesting misclassifications. This is especially appropriate in the case that a model classifies a large number of classes and a misclassification may result in a similar class being predicted, for example the classification of one breed of dog over another in ImageNet \cite{ILSVRC15} . In the paper they try to push the classification of the least likely class ($y_{LL}$) according to the trained model. They name this approach ‘iterative least-likely class method’. They show that this attempt can very effectively break the truthful classification even at a relatively small $\epsilon$ value. This is formalised in a similar way but instead takes steps in the direction of $sign \texttt{\{} \Delta_\mathbf{x}  log   p(y_{LL}|\mathbf{X}\texttt{\}}$, in order to maximise, $log p(y_{LL}|\mathbf{X})$, the probability of the desired output class $y$. In a NN which uses a cross-entropy loss function, this looks like:

\begin{equation}\label{IFGSM}
    \mathbf{x^*}_{N+1} = \mathnormal{Clip_{X,\epsilon}} \texttt{\{} \mathbf{x}_N - \alpha \cdot sign(\Delta_\mathbf{x} \mathbf{J}(\mathbf{w,x},y_{LL})) \texttt{\}}
\end{equation}

The basic iterative method was further built on to incorporate the idea of momentum, a technique used to accelerate gradient descent-based algorithms. This approach is known as Momentum-Iterative FGSM \cite{dong2018boosting}. It works by holding the gradients of the previous $t$ iterations which then contribute to the next step by an amount dictated by the decay factor $\mu$. For example, when $\mu = 0$, the algorithm reduces into iterative FGSM and when $\mu = 1$, all previously saved gradients are added to the current update. This gradient update can be formalised as: 

\begin{equation}\label{MIFGSM1}
    \mathbf{g}_{t+1} = \mu \cdot \mathbf{g}_{t} + \frac{\Delta_\mathbf{x}\mathbf{J}(\mathbf{\mathbf{x^*}_{t+1}},\mathbf{y})}{||\Delta_\mathbf{x}\mathbf{J}(\mathbf{\mathbf{x^*}_{t},y})||_{1}}
\end{equation}

This can then be used in a similar way as the previous methods to generate the next iterative step in the following way: 

\begin{equation}
\label{MIFGSM2}
    \mathbf{x^*}_{t+1} =  \mathbf{x^*}_t + \alpha \cdot sign(\mathbf{g}_{t + 1})
\end{equation}

The use of momentum allows the algorithm to stabilise the updates and navigate through poor local maxima to obtain a method which generally outperforms FGSM and BIM. 

\subsection{Jacobian-based Saliency Map Attack}
\label{subsubsec: JSMA}

The Jacobian-based Saliency Map Attack (JSMA) is one of the most widely recognised and used attack techniques which allows for the crafting of targeted, class specific adversarial samples. It is an example of a white box attack wherein adversaries have full knowledge of network architectures as well as training data. Firstly introduced in \cite{papernot2016limitations}, the proposed attack uses the forward derivatives of a deep neural network to compute a saliency map. The saliency map is then used to traverse through the input space and craft adversarial samples with a target class in mind. Given an input sample $\mathbf{X}$ classified as $\mathbf{F}(\mathbf{X}) = Y$, the adversary attemps to modify the input $\mathbf{X}$ to create an adversarial sample $\mathbf{X^*}$ such that $\mathbf{F}(\mathbf{X^*}) \neq Y = \mathbf{Y^*}$. The adversary looks for the smallest perturbation $\delta_{\mathbf{X}}$ to create an adversarial sample $\mathbf{X^*}$, thereby simultaneously avoiding human detection and satisfying the problem below:

\begin{equation}\label{adversarialgoal}
\argminD_{\delta_{\mathbf{X}}} \norm{\delta_{\mathbf{X}}} \ \mathbf{s.t}\ \mathbf{F}(\mathbf{X} + \delta_{\mathbf{X}}) = \mathbf{F}(\mathbf{X^*}) = \mathbf{Y^*}\\
\end{equation}

The first step required in conducting a successful JSMA is taking the forward derivative of the network with respect to the input sample. The Jacobian matrix for a given input sample $\mathbf{X}$ is computed and is given by:

\begin{equation}\label{jacobian}
    \mathnormal{J_{\mathbf{F}}(\mathbf{X})=\frac{\partial \mathbf{F}(\mathbf{X})}{\partial \mathbf{X}} = \left[ \frac{\partial \mathbf{F}_{j}(\mathbf{X})}{\partial x_i}\right]_{i\in1..M, j\in1..N}}
\end{equation}

Computing the Jacobian allows us to see how small variations in the input space lead to large variations in the output of the network. A large forward derivative for the $\emph{i}^{th}$ feature in a given sample $\mathbf{X}$, indicates that changes to this feature will likely result in large changes to the output of the network. Conversely, features with smaller forward derivative values are less susceptible to adversarial manipulations, because a small derivative value indicates that the feature has a small effect on the output of the network. Thus, computing forward derivatives provides an efficient method to search the input space and craft adversarial samples thenceforth.

The use of saliency maps was first introduced in \cite{simonyan2013deep} as a technique to identify the degree to which pixels in images influence the model to classify the image to a particular class. The aim was to visualize the manner in which classification models use features/pixels to classify a given image. The saliency map was adapted and used to make adversarial saliency maps which allow an attacker to modify the most dominant pixels in an image hence resulting in a misclassification.

The adversary’s goal is to highlight features from an input image $\mathbf{X}$ that cause the network to classify it to a target class $t = \hat{y}(x) = \argminD_{j} \mathnormal{\mathbf{F}_{j}({\mathbf{X}})}$. The output probabilities for the sample $\mathbf{X}$ of target class must be increased while simultaneously decreasing the probabilities of other classes. The adversarial saliency map introduced in \cite{papernot2016limitations} elegantly addresses this and is as follows:

\begin{equation}\label{saliencymap}
    \mathnormal{S(\mathbf{X},t)[i]=}
    \begin{cases}
      0,~ \text{if}\ J_{it}(\mathbf{X})\ < 0 \ \text{or} \sum_{j \neq t} J_{ij} (\mathbf{X}) > 0 \\
       J_{it}(\mathbf{X}) \cdot \abs{\sum_{j \neq t} J_{ij}(\mathbf{X})}, \ \text{otherwise}
    \end{cases}
\end{equation}

where $J_{ij} (\mathbf{X})$ is $J_{\mathbf{F}}[i,j](\mathbf{X}) = \frac{\partial \mathbf{F}_{j}(\mathbf{X})}{\partial \mathbf{X}_i}$.

Put simply, the first condition in the first line ignores any features that are negatively correlated with the target class. Alternatively, the second condition on the first line ignores features that are positively correlated with classes that are not the target class $\mathnormal{j\neq{t}}$. This second condition does however allow for negative or constant gradients for classes $\mathnormal{j\neq{t}}$. In this way features which decrease the probability of all other classes are chosen as well. The product term on the second line considers features that both increase the probabilities of the target classes, as well as decrease the probabilities of all other classes. It follows that high values of $\mathnormal{S(\mathbf{X},t)[i]}$ correlate with features that increase the likelihood of the model’s classification to the target class $t$. This version of the saliency map highlights features the adversary should increase to result in a misclassification. Conversely, a complementary version of the saliency map detailing the features an adversary should decrease to achieve a misclassification can be seen as follows:

\begin{equation}\label{saliencymap2}
    \mathnormal{\Tilde{S}(\mathbf{X},t)[i]=}
    \begin{cases}
      0,~ \text{if}\ J_{it}(\mathbf{X})\ > 0 \ \text{or} \sum_{j \neq t} J_{ij} (\mathbf{X}) < 0 \\
       \abs{J_{it}(\mathbf{X})} \cdot \bigg(\sum_{j \neq t} J_{ij}(\mathbf{X}) \bigg), \ \text{otherwise}
    \end{cases}
\end{equation}

Once the saliency map is computed, the input space must be traversed and perturbed to satisfy the equation in (\ref{adversarialgoal}). Researchers from \cite{papernot2016limitations} approach this by introducing an algorithm which iteratively
modifies a sample $\mathbf{X}$ by considering two pixels at a time. The algorithm conducts the search across the input domain $\Gamma$ over all input indices \cite{wiyatno2018maximal}. These pixels are selected by the saliency map, they are then perturbed, and removed from the input search domain. This process repeats until the sample is classified by the neural network as the target class, or until a criterion $\Upsilon$, which specifies the maximum number of iterations, is reached. The criterion $\Upsilon$ tries to put an upper limit on the number of perturbed features so that the adversarial sample remains undetectable by humans.

Results from \cite{papernot2016limitations} where researchers conducted the JSMA attack on the LeNet architecture trained on the MNIST dataset, show that any input digit from a source class can be perturbed to fit any other class specified by the adversary with a 97.1\% success rate. They also maintain that the adversary only needs to change an average of 4.02\% of input features per sample. They also conducted experiments to correlate the percentage of distorted pixels for a given digit, to the humans' ability to recognize the digit. As a result of the experiment, they identified the threshold of 14.29\% wherein images which had a greater feature distortion than this, were unrecognizable by humans.

\subsection{Generative Model Based Method}
\label{subsubsec:GANs}

Developed by \textcite{goodfellow2014generative} in 2014, a Generative Adversarial Network (GAN) is a generative model which solved the data generation problem without the use of Markov Chains. A GAN is comprised of two neural network architectures: a generator, and a discriminator which is sometimes called a critic. These two components act in opposition to one another and converge to a solution wherein the generator is able to produce data from the underlying distribution of input samples $\mathbf{x}$.

Given input samples $\mathbf{x}$, the generator maps noise variables $z$ into data space to try to produce samples $\mathcal{G}(z)$ that originate from the input samples underlying distribution $\mathcal{P(\mathbf{x})}$. The discriminator on the other hand tries to discern whether the samples come from $\mathbf{x}$ or are from the generator. $\mathcal{C}(x)$ is the output of the discriminator and is the probability that $x$ came from the input data and not from the generator. The training of the GAN therefore involves simultaneously maximising $\mathcal{C}(x)$ and minimising $\log(1-\mathcal{C(G}(\mathbf{z})))$. The loss function can be described as follows:

\begin{equation}\label{GANloss}
\mathcal{\min_{G} \max_{C}} V(\mathcal{C,G}) = \mathbb{E}_{x\sim p_{x}(\mathbf{x})}[\mathcal{\log C(\mathbf{x})}] + \mathbb{E}_{z\sim p_{z}(z)}[\log(1 - \mathcal{C(G}(z)))].
\end{equation}

A clever implementation of a GAN to construct adversarial examples was demonstrated by researchers in \cite{zhao2017generating}. The aim of the research was to craft more natural looking adversarial samples in contrast to those generated by already existing adversarial attack methods such as the FGSM. For a given input, they search for adversarial examples in the local region of its analogous representation in a low-dimensional space. To do this, they define the $z$ space. The $z$ vector space is a dense latent low-dimensional representation of input samples from domain $\mathbf{X}$. Instead of finding an adversarial sample $x^*$ directly, they find an adversarial $z^*$, such that mapping this to the input space produces an adversarial sample $x^*$. 

The training architecture used consists of GAN and an additional component which they call an inverter. The inverter acts in the opposite manner in which a generator in a GAN works. It takes a sample $x$ and maps it into its analogous dense representation in $z$ space $z^{’} = \mathcal{I}_{\gamma}(x)$. Due to the unstable nature of GANs in their original form, they used a modified loss function introduced by \cite{ arjovsky2017towards}. The loss function is called the Wasserstein loss and it has been proven to help the stabilization of GANs. The loss function is as follows:

\begin{equation}\label{wassersteinloss}
\min_{\theta} \max_{\omega} \mathbb{E}_{x\sim p_{x}(x)}[\mathcal{C}_{\omega}(x)] - \mathbb{E}_{z\sim p_{z}(z)}[\mathcal{C}_{\omega}(\mathcal{G}_{\theta}(z))].
\end{equation}

\begin{figure}[ht]
  \centering
	\includegraphics[scale=0.55]{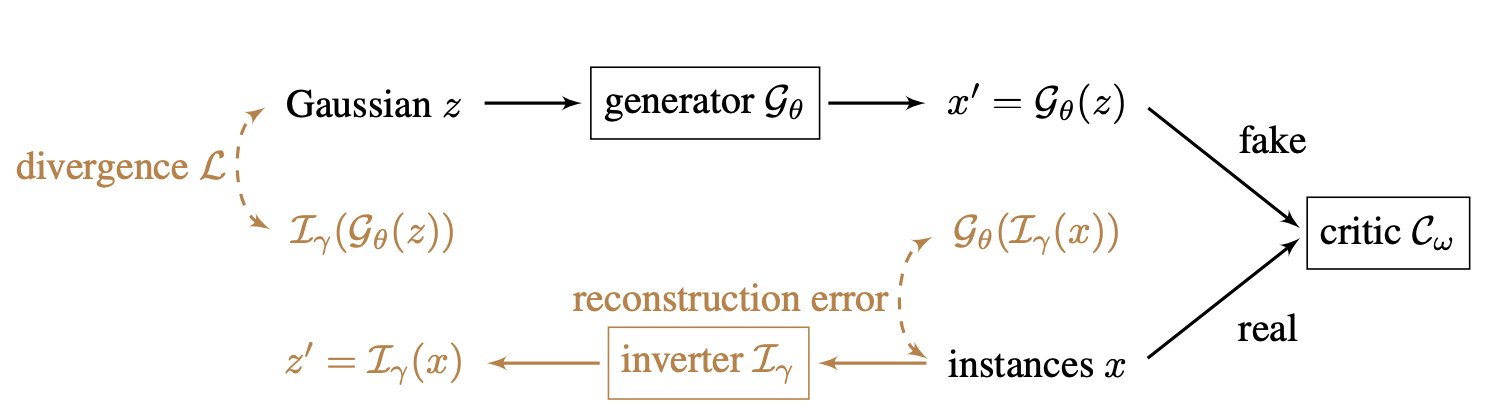}
    {\small\flushleft Source: \textcite{zhao2017generating}}
  \caption{The training procedure of the GAN and the Inverter. The diagram illustrates the components that are compared to build the loss function of the inverter}   
    \label{fig:GANInveter}
\end{figure}

They firstly train the WGAN until its generator is able to map random noise vectors to samples that come from domain $\mathbf{X}$, then train the inverter $\mathcal{I}$ until it is sufficiently able to produce a low dimensional representation of samples from domain $\mathbf{X}$. To train the inverter, they define two error terms: the \emph{reconstruction} error, and the \emph{divergence} error. In order to understand these definitions we need to understand the inputs and outputs of parts of the model. The output of the inverter $\mathcal{I}_{\gamma}(x)$ is fed into the trained generator to get $\mathcal{G_{\theta}(I_{\gamma}}(x))$. The difference between this output and the sample $x$ is called the reconstruction error. The generator on the other hand gives the output $\mathcal{G}_{\theta}(z)$, for a noise vector $z$. This is fed into the inverter to produce $\mathcal{I}_{\gamma}(\mathcal{G}_{\theta}(z))$. The difference between this value and the sampled $z$ is called the divergence error. A representation of their architecture is demonstrated in Figure \ref{fig:GANInveter}. The loss function used to train the inverter aims to minimise the summation between these two values, shown in Equation \ref{inverterloss}:

\begin{equation}\label{inverterloss}
\min_{\gamma} \mathbb{E}_{x\sim p_{x}(x)}  \norm{\mathcal{G}_{\theta}(\mathcal{I}_{\gamma}(x)) - x} + \lambda \cdot \mathbb{E}_{z\sim p_{z}(z)}[\mathcal{L}(z,\mathcal{I}_{\gamma}(\mathcal{G}_{\theta}(z)))]
\end{equation}

They researchers in \cite{zhao2017generating} search for adversarial samples using an iterative stochastic search and a hybrid shrinking search. The former starts the search in a small region then incrementally expands the search region until an adversarial sample is found. The latter does the opposite, it begins with a wide search range then repeatedly narrows it down until an adversarial sample is reached. This algorithm proved to be 4x more computationally efficient than the aforementioned algorithm. They tested their adversarial sample generation technique to generate text and image adversaries. They show that using this technique, they were able to generate natural looking, and more legible, adversarial examples.

\subsection{Universal Attacks}
\label{subsec:universal attacks}

Universal attacks are a special type of attack used to generate a single, fixed, universal perturbation image, which, when added to any natural image, causes it to be misclassified with a high probability. Simply put, a single perturbation vector is generated with the ability to fool the target model on most natural images \cite{moosavidezfooli2017universal}. These universal perturbations have also been shown to work very well, managing to fool multiple state of the art classification classifiers, with fooling rates of up to 93 \%. It has also been shown that some universal perturbations computed for specific networks, generalise well and perform relatively well on completely different architectures, and so are described in \cite{moosavidezfooli2017universal} as ‘to some extent, doubly universal’. For example, a universal perturbation fashioned for a VGG-16 model still managed to a minimum fool rate of 56 \% across the other 5 models in the study. 

\begin{figure}[ht]
  \centering
	\includegraphics[scale=0.4]{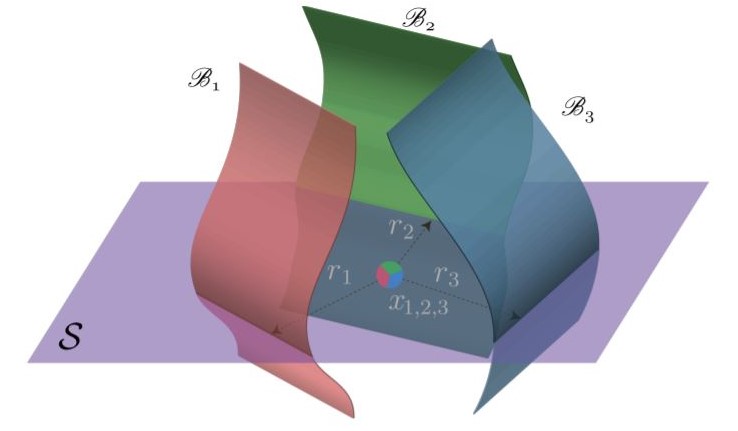}
    {\small\flushleft Source:  \textcite{moosavidezfooli2017universal} }
  \caption{Illustration of a trivial example of lower dimensional subspace S containing normal vectors to the decision boundary. It displays the superimposition of three datapoints ${xi}{i=1}^3$, and the adversarial examples, ${ri}{i=1}^3$ that send the respective points to the decision boundary ${Bi}{i=1}^3$. It is important to note that all ${ri}{i=1}^3$lie within the subspace S.}   
    \label{fig:universalperturbationlowerdim}
\end{figure}

The existence of successful universal perturbations highlights an inherent vulnerability in deep neural networks. The decision boundaries of these networks are highly dimensional and therefore contain redundancies and large “geometric correlations between different parts of the decision boundary” \cite{moosavidezfooli2017universal}. This indicates the existence of a lower dimensional subspace which captures these correlations by collecting normals to different regions of the decision boundary. This is illustrated in figure \ref{fig:universalperturbationlowerdim}. This can be exploited as perturbations in this subspace are thus likely to lead to datapoints in this region fooling the model.

\section{Countermeasures}
\label{sec:countermeasures}

Whilst adversarial attacks can be powerful, there are a number of countermeasures which help to defend against or even mitigate them. These approaches can either be proactive or reactive. The goal of proactive approaches is to strengthen the model itself against adversarial attacks; in essence making it more robust. Alternatively, reactive approaches aim to detect adversarial attacks and react appropriately, completely regardless of the model \cite{8611298} \cite{ REN2020346} \cite{ deng2020analysis}.

Common proactive approaches include adversarial training and defensive distillation. Adversarial training involves retraining the model on a dataset including adversarial examples \cite{goodfellow2015explaining}, while defensive distillation trains a new model using the outputted class probabilities of the original model as training labels \cite{papernot2016distillation}. 

Denoising is another class of proactive approaches which, as the name suggests, look to reduce noise of the input to highlight and clean the effect of adversarial tampering. Feature squeezing is the primary example of this and works by reducing the dimensions of the input, thereby reducing the search-space available for an adversary to tamper. Adversarial examples can be identified by comparing the output of the model given both the original and squeezed input \cite{featuresqueeze}. Singular Value Decomposition can also be used similarly to denoise input data \cite{svd}. 

Reactive approaches include adversarial detection or input reconstruction. Adversarial detection generally works by utilising a separate mechanism for determining the legitimacy of input data. There have been a number of detection methods proposed with some success. It should be noted that while some approaches naturally lead to a suitable fix \cite{ song2018pixeldefend}, some do not and thus would probably deal with the attack in an appropriate manner such as to ignore the adversarial data completely \cite{lu2017safetynet}. Input reconstruction works to transform adversarial examples back to their original form, while normal, clean examples are left completely or minimally changed.

\subsection{Proactive}
\label{subsec:proactive}

\subsubsection{Adversarial Training}  \hspace{0pt} \par
\label{subsubsubsec:adversarialtraining}
\noindent Adversarial training is a method of defending against adversarial examples that is intuitive and commonly accepted as one of the most successful techniques for improving model robustness when put into practice. It reinforces the neural network through the proactive inclusion of adversarial samples in the training data and can be seen through the perspective of a robust optimisation for a min-max problem \cite{advtrain}. The concept behind adversarial training is most simply formulated by \textcite{defences}, representing the adversarial loss as $J(\theta, x', y)$, in which the network weights are $\theta$, adversarial inputs are $x'$ and the ground truth label is $y$:

\begin{equation}\label{eq:advtrain}
\min_{\theta}\max_{D(x, x')<\eta} J(\theta, x', y)
\end{equation}

The term on the left in the maximisation constraint $D(x, x')<\eta$ of Equation \ref{eq:advtrain} denotes a metric for the distance between the original input $x$ and adversarial example $x'$. This maximisation is used to generate an adversarial example that is the most effective and hence has the largest value for the distance metric. The method for generating the example can vary, such as Fast Gradient Sign Method (detailed in Section \ref{subsubsubsec:fgsm} on attacks) and Projected Gradient Decent.

The minimisation term outside of the maximisation is the standard neural network procedure for training by minimising the loss through fine-tuning of the network’s weights. The result is a network robust to the adversarial attack used for generating the adversarial example in the training stage and can achieve state-of-the-art accuracies on many competitive benchmarks. Earlier implementations of adversarial trained models estimated the loss function linearly and hence were unprotected against and vulnerable to iterative adversarial generation methods; recent advances in adversarial training are resistant to even iterative attacks \cite{advtrain}.

\subsubsection{Defensive Distillation}  \hspace{0pt} \par
\label{subsubsubsec:distillation}
\noindent Defensive distillation is a process that uses knowledge distilled from a neural network architecture in order to increase its robustness to adversarial examples \cite{papernot2016distillation}. This idea was extended from work done in \cite{hinton2015distilling}, where they reasoned that computational complexity of larger architectures could be bypassed by using smaller architectures that contain knowledge of larger, previously trained neural networks. The intuition behind this is that knowledge from complex architectures can be fitted onto smaller resource-constrained devices \cite{papernot2016distillation}.

\begin{figure}[ht]
  \centering
	\includegraphics[scale=0.4]{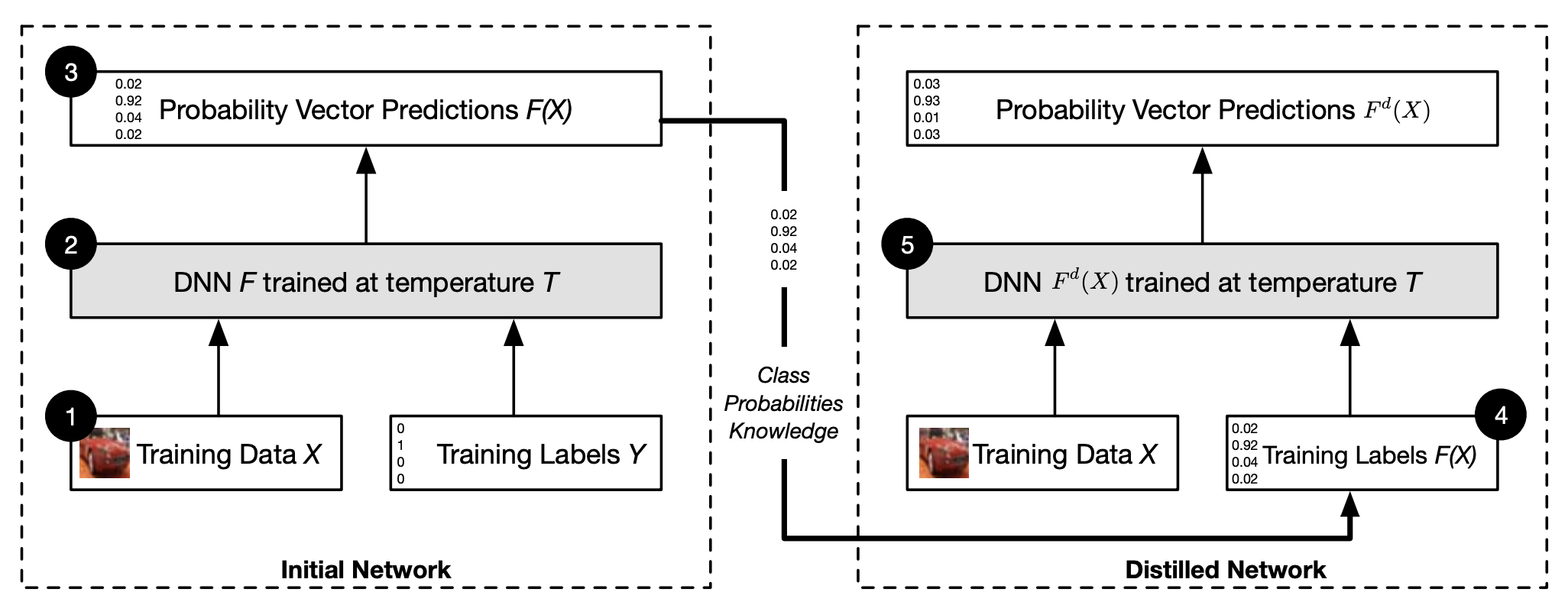}
    {\small\flushleft Source: \textcite{papernot2016distillation}}
  \caption{The training produce of a distilled network as proposed by \cite{papernot2016distillation}. The network on the left is initial network which outputs probability vector prediction. The figure on the left uses these probability targets as labels to train the distilled network.}   
    \label{fig:distillationdiagram}
\end{figure}

The distillation training procedure involves the utilisation of ‘soft labels’ output by the first neural network, to train a second neural network. The only difference is in the training procedure of soft labels. \textcite{papernot2016distillation} argue that the use of soft targets in training is justified because the probabilities encode extra knowledge about each class. The vector of soft labels that are output for each datapoint, represents the probability of membership to each of the classes, with the highest probability being the most likely class the datapoint belongs to. The soft labels for a datapoint are calculated in (\ref{softmax_eqn}),

\begin{equation}\label{softmax_eqn}
qi = \frac{\exp(z_i/T)}{\sum_{j} \exp(z_j/T)}
\end{equation}

where \emph{$q_i$} is the class probability for the \emph{i-th} class, \emph{$z_i$} the input to the softmax layer and \emph{T} the temperature, which controls how 'soft' the probability is distributed over the classes \cite{hinton2015distilling}. The higher the temperature value used, the more the output probability distribution resembles a uniform one. A breakdown of the distillation procedure is illustrated in Figure \ref{fig:distillationdiagram}. 

\textcite{papernot2016distillation} built a distilled network using the same architecture which provided the soft labels. They used the same high-temperature values to train both networks, then set the temperature back to 1 during test time. They trained, tested, and measured the model’s sensitivity to adversarial examples when trained on the MNIST dataset and the CIFAR10 dataset. They crafted adversarial samples using a succession of two techniques: direction sensitivity estimation, and perturbation selection.

A reduction in adversarial sample success rate from 95.89\% to 0.45\% was noticed in the model trained on the MNIST dataset and a drop from 87.89\% to 5.11\% on the model trained in the CIFAR10 dataset. The percentage changes observed were from a model with no distillation to a distilled model, both trained with temperature values of 100. Conflicting results are obtained in \cite{soll2019evaluating} where a distilled network was used in defending text processing neural networks. Little to no change in the robustness of their network was obtained when a distilled network was used. They speculate that the use of distillation in text processing is misplaced, as gradients are not directly added to the input, but are rather just used to highlight the importance of a given word.

The distillation method seems to succumb to the black box attack proposed in \cite{papernot2017practical}. The researchers recreated the models trained in \cite{papernot2016distillation} and successfully implemented an FGSM attack under black box conditions. They accredit the success of the attack to a phenomenon called \emph{gradient masking}, the method in which distillation defence works. They postulate that the distillation method makes the gradient harder for the adversary to obtain, but does not nullify the attack vector itself.

An extension to the defensive distillation was made by co-authors of the original papers of \cite{papernot2016distillation}. In the revised paper \cite{papernot2017extending}, they attempt to bypass the gradient masking effect by making modifications to the original defensive distillation process. They add two major additional components to their new model: an uncertainty metric for the inputs to the network, and an outlier class with the sole purpose of classifying adversarial samples. These are accompanied by changes made to the training procedure of networks, namely a modification to the loss function in the distilled network, and the setting of the temperature value to 1 at all times. The result of this was a network that proved to be less likely to suffer from gradient masking. Because of this, black box attacks were more likely to be flagged by the model as outliers, as supposed to white-box attacks. They conclude by stating the following, ‘Defensive distillation’s most appealing aspect remains that it does not require that the defender generate adversarial samples’ \cite{papernot2017extending}.

\subsubsection{Denoising}  \hspace{0pt} \par
\label{subsubsubsec:distillation}

\noindent Denoising is a class of proactive defensive measures to mitigate the efficacy of adversarial attacks by removing or reducing the perturbations introduced in the adversarial example. Denoising can be categorised into two subsections, dependent on the stage of the model where the denoising occurs. One direction opts to cleanse the original input, before passing it to the model, while the other sterilises the extracted features learnt through the NN algorithm. \cite{defences}

\subsubsection{Feature Squeezing}  \hspace{0pt} \par
\label{subsubsubsec:distillation}

\noindent Feature squeezing is an accurate and cost-effective adversarial countermeasure, developed on the fundamental principle that a high frequency of the input spaces of features are significantly larger than necessary. Considerable opportunities to generate adversarial examples are afforded to an adversary due to the expansive input space. This defensive measure limits the number of input features by computing and removing any that are unnecessary, greatly decreasing the freedom of the adversary to generate examples \cite{featuresqueeze}. The method inherits its name due to this “squeezing”. 
Adversarial detection is trivial after the squeezing, the outputs of the NN model for both the original input and squeezed input are compared and if there are outstanding differences then the input can be labelled as adversarial. The exact level of contrariety at which the determination of an adversary occurs can be controlled through a threshold parameter. Legitimate inputs detected will have their model predictions outputted, however those of adversarial inputs can be discarded \cite{featuresqueeze}. Although feature squeezing’s primary application is image classification, the method is transferable to other domains.

\begin{figure}[ht]
  \centering
	\includegraphics[scale=0.5]{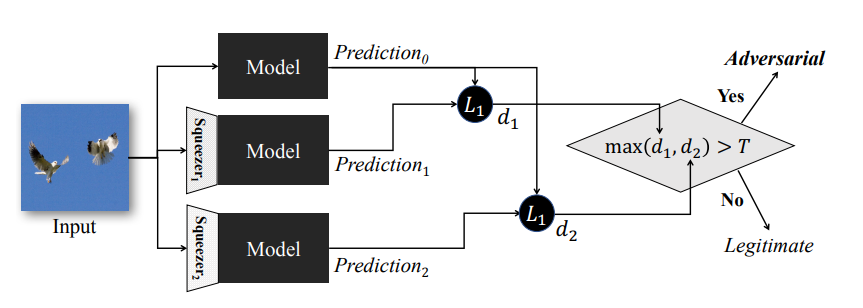}
    {\small\flushleft Source: \textcite{featuresqueeze}}
  \caption{The methodology proposed by \textcite{featuresqueeze}. The model’s prediction on a squeezed and original input image are compared at 'd1' and 'd2'; the threshold used to classify adversarial examples is denoted as 'T'.}   
    \label{fig:featuresqueezediagram}
\end{figure} 

The feature squeezing proposed in \textcite{featuresqueeze} is achieved through two straightforward methods. The first approach is the reduction of each pixel’s colour depth within an image. Secondly, differences between the pixels are diluted via spatial smoothing. Feature squeezing through these methods have been demonstrated to considerably strengthen the robustness of the predictions from a neural network model. Legitimate (non-adversarial) inputs are left unaltered and hence maintain the same accuracy; static (non-adaptive) adversarial examples have correct labels predicted using the squeezed input \cite{featuresqueeze}.

Neural networks require an assumption of a continuous input space due to their nature as differentiable models. This is contrary to a digital device’s restriction to approximate natural continuous data to a discrete calculation. Matrices of pixels constitute digital images, each pixel being a number value representing a specified colour. Colour bit depths in popular image representations can often lead to features that are not relevant and hence their reduction can maintain the predictive capability of the model while reducing the freedom of an adversary to generate an attack. One of the popular representations of images is 8-bit greyscale, affording $2^8 = 256$ possible values, in this case ranging from white at 0, through various shades of grey, to black at 255. An extension of this 8-bit representation uses three separate colour channels: red, green, and blue, each with their own respective 8-bits, to create a 24-bit colour image, producing $2^{24} > 16\ million$ different possible colours for each pixel. The human eye is only able to distinguish around ten million colours \cite{colourdepth}, hence why this representation is often referred to as ‘True Colour’. Larger bit colour depths are often preferred by people as the image produced is a more natural representation and closer to the truth, however they are not necessary for understanding and recognising an image, hence why black-and-white images are interpretable.

The implementation of bit depth reduction in \textcite{featuresqueeze} takes an input and produces an output on an identical numerical scale $[0, 1]$, therefore the NN model requires no alteration. To reduce an image to $i$-bit depth, such that $1 \leq i \leq 7$, the product of the input value and $2i – 1$ is taken and subsequently rounded and re-scaled to $[0, 1]$ before being divided by $2i – 1$. The process of rounding to integers during this method causes the reduction in information capacity from an 8-bit to $i$-bit representation \cite{featuresqueeze}.

\begin{figure}[ht]
  \centering
	\includegraphics[scale=0.5]{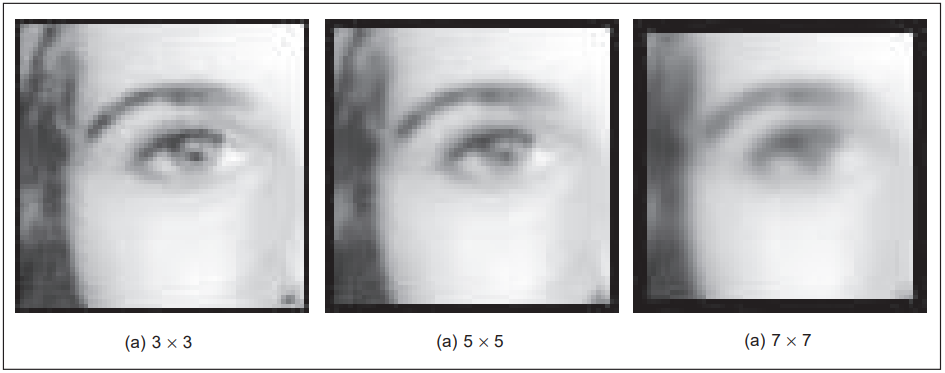}
    {\small\flushleft Source: \textcite{smoothing}}
  \caption{Application of Gaussian local smoothing on an image, with kernel size of a) $3\times3$ , b) $5\times5$ , c) $7\times7$. Increasing levels of zero padding are required, and visible in the thicker black borders around the image.}   
    \label{fig:smoothingdiagram}
\end{figure} 

Spatial smoothing is a common image processing procedure for reducing the amount of noise within the image and is also known as blur. The smoothing applied in the feature squeezing proposed by \textcite{featuresqueeze} is categorised into to variants, local and non-local smoothing.

Local smoothing is a smoothing mechanism that makes use of the neighbouring pixels within a window (also known as a kernel) around a target pixel, assigning different weights to them depending on the exact smoothing method used, such as Gaussian, mean or median smoothing \cite{smoothing}. In \textcite{featuresqueeze}, median smoothing is used, a window slides over the image in a raster scan fashion, replacing the central pixel value with the median of all pixels in the window. Depending on the use of padding, this method does not reduce the size of the image and smooths out values across pixels that are nearby. Adjacent pixels become more similar in value and this process can cause features to be squeezed out. Smoothing methods such as median smoothing have variable parameters that are configurable, such as the window size which can be set from 1 pixel at a minimum, to the dimensions of the whole image at a maximum. For feature squeezing with median smoothing, a square kernel of size 2x2 with reflection padding at the edges is used. A window of this shape takes the bottom right pixel as the target pixel and biases towards the larger value when calculating the median from an even set of numbers. The padding method used mirrors the image along the edges for calculating the values of pixels at the extremities, as there would not be pixels to fill the window. This smoothing is commonly known for being highly practical for removing salt-and-pepper noise from images, random and sparse occurrences of black and white pixels, while being able to maintain defined edges of objects within the image \cite{featuresqueeze}\cite{smoothing}.
By contrast, non-local smoothing applies its smoothing methods over a collection of pixels that extends much further than a small, localised window. For a target image patch, non-local smoothing will calculate other patches that are similar across an area of the image and replace the target patch with the computed average of the patches. Due to the random and sparse distribution of noise within an image, it is assumed that the mean of the noise will be zero and hence this method is able to eliminate noise from the image while leaving edges unaffected. The mechanism used to assign weights to similar patches during the averaging closely resembles local smoothing: Gaussian, mean and median operators are commonly used \cite{nonlocal}. \textcite{featuresqueeze} use a popular adaptation of a Gaussian kernel. This enables additional control of the deviation from the mean, as well as the window size across which similar patches are searched for, the patch size and a variable for controlling the strength at which the filter is applied.

\subsubsection{SVD}  \hspace{0pt} \par
\label{subsubsubsec:distillation}

\noindent In adversarial attacks against pattern recognition systems, the adversary introduces specially crafted perturbations to the pixel values of the target image to generate what are known as adversarial examples. As high-dimensional space is linear in nature, even a slight pixel change can have a large impact and be dramatically increased in the feature space, ultimately misleading the neural network model to misclassify an image with a potentially high confidence. Singular Value Decomposition is a commonly used technique when analysing multivariate data \cite{svdpca}. Research has concluded that the SVD of an image can also be used as an adversarial defence by computing an optimal approximation of the image matrix in terms of square loss, as a solution for eliminating the perturbations caused by the adversary \cite{svd}.

In order to robustly defend against an adversarial attack, it is necessary for the humanly imperceptible perturbations introduced to the input before the input is passed onto the target model. The difference between the unaltered, original input, $X$, and the adversarial example, $X^{ADV}$, is calculated as the minute value $\eta$:

\begin{equation}\label{eq:svdadv}
||X^{ADV} - X || = \eta
\end{equation}

A matrix’s distribution characteristics are described by the unique singular values (principal components) $\delta_1,\delta_2,\ldots,\delta_r,\forall A \in C^{m*n}$. A linear transformation of the matrix $A$ maps its points in $m$-dimensional to $n$-dimensional space \cite{svd}. The SVD of a matrix decomposes the transformation into three separate matrix parts, $U$, $s$ and $V$. $U$ and $V$ are complex unitary matrices whereas $v$ is a diagonal matrix. Lossy compression of an image can be achieved through an orthogonal transformation, which retains defining features of the image but with reduced data. This is common for a variety of techniques and processes such as the storage, transmission and analysis of images. A digital image $A$ can be interpreted as having a vertical and horizontal time-frequency in a two-dimensional representation, upon which the singular value decomposition can be computed \cite{svdpca}\cite{svd}:

\begin{equation}\label{eq:svddecomp}
A = UsV^\top = U \begin{bmatrix} 
									\delta \& 0 \\
									0 \& 0
									\end{bmatrix}
                                    V^\top = \sum_{i=1}^{j} \delta_iu_iv_i^\top
\end{equation}
 
\begin{figure}[ht]
  \centering
	\includegraphics[width=\textwidth]{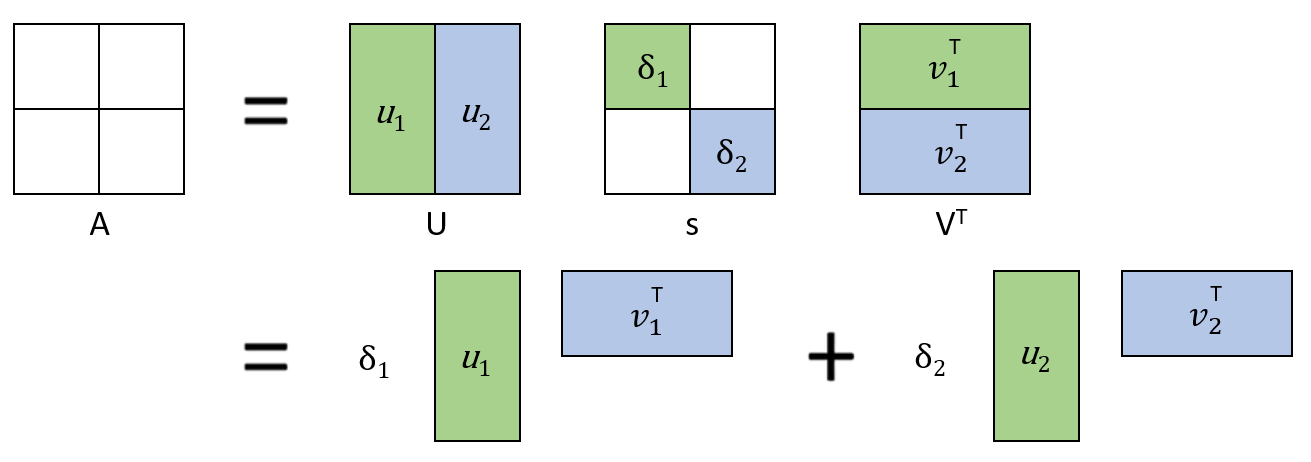}
  \caption{Diagram based on a similar figure by \textcite{svd}, providing a detailed and concise breakdown of the SVD process.}   
    \label{fig:svddiagram}
\end{figure}

In Figure \ref{fig:svddiagram}, a concise visualisation of the singular value decomposition of a $2 \times 2$ matrix is displayed. In this figure and Equation \ref{eq:svddecomp}, matrix $U$ and $V$ have the respective column vectors $u_i$ and $v_i$ and $\delta_i$ is a non-zero and real singular value from the matrix $s$. A summation over $j$ many sub-graphs, formed from the product of the corresponding column vectors and values from the decomposition, will reform the original image. Ordering these sub-graphs by singular value and only including a specified number of the largest will preserve the most information in the image possible. The information about the geometry and texture of the image are predominantly represented in matrices $V$ and $U$, whilst $s$ stores the energy information in the singular values \cite{svd}. Singular values have three distinct and provable properties: singular values are stable, singular values are proportional invariant, singular values have rotation invariance and matrix approximation can be optimally achieved by selecting singular values of matrix $A$ greater than a certain threshold. This final property is essential for data compression via matrix reduction and the principal justification for the use of SVD as a defensive method to remove the imperceptible perturbations in an adversarial example whilst maintaining all other core information.

\subsection{Reactive}
\label{subsec:reactive}

\subsubsection{Adversarial Detection}  \hspace{0pt} \par
\label{subsubsubsec:adversarialtraining}

\noindent Adversarial detectors can take many forms. One of the most basic adversarial detection approaches are DNN-based binary classifiers to estimate the probability an input is adversarial. \cite{metzen2017detecting} created exactly this and were able to show that a relatively small NN can act as a subnetwork to detect adversarial examples “surprisingly well”. A slightly more robust example is that used in SafetyNet \cite{lu2017safetynet}, which is able to detect, with high accuracy, adversarial examples built from a number of attacking approaches. These detected adversarial examples are rejected by the model based on their classification confidence. SafetyNet also is shown to be relatively robust to type II attacks which look not only to fool the main classifier into mislabelling the input, but also to evade the detector. The system works by taking the binary threshold of each ReLU layer’s output as features to a radial basis function (RBF)-SVM classifier, which then detects adversarial examples. They argue this works well even when the adversary is aware of the detector. This is because it forces the adversary to solve the difficult optimisation problem of discovering an optimal value for both creating adversarial examples, as well as optimising the new features of the adversarial detector. \\

Similarly, \cite{feinman2017detecting} introduced two new effective indicators of adversarial examples, which work well together in complementary situations: Bayesian neural networks and kernel density estimation. The kernel density estimation approach works by using the features on the final hidden layer of the NN to build a picture of the submanifolds of each class. From this we can estimate the distance a point is from the submanifold of the appropriate target. This approach works well for adversarial examples far from the submanifold of the target, however, may not when closer. In this case, they show that Bayesian neural networks can be used to distinguish adversarial examples by estimating the uncertainty of inputs. This works because they were able to show that putting adversarial examples through Baysian models, such as the Gaussian process \cite{10.5555/1162254}, tends to produce higher levels of uncertainty.\\

A different approach, used as part of PixelDefend \cite{song2018pixeldefend}, an input reconstruction approach, looks at the training distribution and where example inputs fall within that. They demonstrated that PixelCNN \cite{NIPS2016_b1301141} \cite{salimans2017pixelcnn}, a state of the art neural density model, can be used to effectively identify perturbations. They showed that the distribution of log-likelihoods highlighted a substantial difference between normal images and adversarial ones and that even small perturbations of 3\% can lead to sizable reductions in the log-likelihood. Along with the permutation test \cite{EfroTibs93} they were then able to calculate the exact p-values. These p-values define the probability that an input is drawn from the training distribution. By examining the generated p-values the adversarial examples are distinguishable as they generally lie in lower probability regions. 

\subsubsection{Input Reconstruction}  \hspace{0pt} \par
\label{subsubsubsec:adversarialtraining}

\noindent The key idea behind input reconstruction is to transform the adversarial example back into clean data before passing it through to the model. A good example of this approach is, PixelDefend \cite{song2018pixeldefend} which reconstructs adversarial images back to the training distribution. As mentioned above, PixelDefend uses PixelCNN as their detecting mechanism. PixelDefend leaves datapoints that are already sufficiently within the distribution of the training data unchanged. Meanwhile, applying small deviations to datapoints far enough outside the training distribution. With the pixelCNN distribution, $p_{CNN}(X)$, used as an approximation of the probability of $x$, $p(X)$, they frame the problem as an optimisation one. The problem is to maximise $p(X^*)$ subject to constraints on the size of the perturbation, $\epsilon_{defend}$. I.e $\max{p(X^*)}$\\

To achieve this, they take a greedy approach not too dissimilar to the generation of images from PixelCNN, with the added constraint that the image should be within the $\epsilon$-neighbourhood. As PixelCNN is an autoregressive model, this process is typically a slow one, so they use a faster version proposed by Ramachandran \cite{ramachandran2017fast}. \\

\begin{figure}[ht]
  \centering
	\includegraphics[scale=0.9]{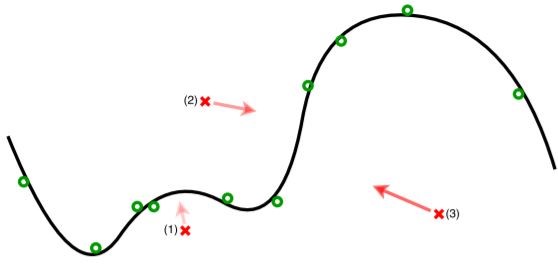}
    {\small\flushleft Source: \textcite{meng2017magnet}}
  \caption{Simple 2-D illustration of how MagNet works. The manifold of normal examples is represtned by the curve. Green circles and red crosses correspond to normal examples and adversarial examples respectively. The transformation by an autoencoder is show by the red arrow. The detector measures reconstruction errors and rejects those with errors too large, e.g. cross 3. Adversarial examples closer to the manifold are reconstructed to the original by the reformer.
}   
    \label{fig:MagNetIllustration}
\end{figure}

A similar example is MagNet \cite{meng2017magnet} which is comprised of two components: a detector, which rejects examples too far from the manifold boundary, and the reformer, which given an input $x$, endeavours to generate an $x’$ as a close approximation of $x$ that falls on or near the manifold. It then gives $x’$ to the target classifier in the place of $x$, thereby reconstructing the adversarial example. This process is illustrated in \ref{fig:MagNetIllustration} \\

In \cite{meng2017magnet}, the detector is realised as a binary classifier, mapping inputs to either adversarial or clean, as discussed above. The implementation opted for is based on probability divergence and is very similar to the kernel density estimation used in \cite{feinman2017detecting}. They make the use of an autoencoder trained to learn the features of the training data, as the detector. An autoencoder is used to encode its input and then decode it in an attempt to recreate the input. A reconstruction error is used to approximate the distance between an input and the learned manifold of clean examples. Now, if an input is drawn from the same data generation process as the training set, the reconstruction error should be low, and thus the reconstruction error can act as distance measure from the manifold of normal, clean data. However as discussed previously, this tends to be less effective when the reconstruction error becomes smaller as adversarial examples are made closer to the original. This can be dealt with by passing the both the inputs and the output of the autoencoder through the model and comparing. If for example, $x$ is clean and drawn from the same process as the training data, the output of the autoencoder $ae(x)$ should be very similar. Therefore, the output of the model should be similar also, i.e $f(x) = f(ae(x))$. However, if $x’$ is an adversarial example, even if the reconstruction error is low, $f(x’)$ and $f(ae(x’))$ can be significantly different. This works by taking advantage of the target classifier, assuming the use of softmax at the last layer of their network, an assumption which is the case in most neural network classifiers. \\

The reformer tries to reconstruct adversarial inputs appropriately, so they are close to the original image, while leaving clean inputs unchanged. Similar to the detector, \cite{meng2017magnet} propose the use of autoencoders in the implementation of the reformer, to take advantage of the distribution of normal data. They create a large number of autoencoders which are all trained to minimise the reconstruction error on the training set. This way, given a clean datapoint, the output should be very similar, but given an adversarial datapoint, the expected output should approximate the adversarial example but be closer to the manifold of the normal examples. With multiple autoencoders, MagNet randomly picks one as each defensive device every time. This forces the adversary to have to train their adversarial examples to work on multiple autoencoders at the same time, assuming they cannot predict which autoencoder is picked, and provides diversity in defence.

\section{Conclusions}
\label{sec:background}
\noindent This work has provided a comprehensive overview of the principles of adversarial machine learning attacks and existing countermeasures. Our analysis has shown that this threat poses a very real danger to machine learning algorithms, which play a massive and growing role in all aspects of modern society, including safety-critical applications and advance defence systems. There remain several outstanding challenges including efficient deployment mitigating techniques that balance the security risks with the expected overheads. This is specially true for resources-constrained systems. This requires the development of specific implementations of these countermeasures tailored to each application scenario.

\section{Acknowledgments }
\label{sec:background}
\noindent This research was partly funded by the royal academy of
engineering (grant No. IF202136).

\cleardoublepage
\phantomsection
\printbibliography

@article{defences,
    title = {Adversarial Attacks and Defenses in Deep Learning},
    journal = {Engineering},
    volume = {6},
    number = {3},
    pages = {346-360},
    year = {2020},
    issn = {2095-8099},
    doi = {https://doi.org/10.1016/j.eng.2019.12.012},
    url = {https://www.sciencedirect.com/science/article/pii/S209580991930503X},
    author = {Kui Ren and Tianhang Zheng and Zhan Qin and Xue Liu},
}

@article{featuresqueeze,
  author    = {Weilin Xu and
               David Evans and
               Yanjun Qi},
  title     = {Feature Squeezing: Detecting Adversarial Examples in Deep Neural Networks},
  journal   = {CoRR},
  volume    = {abs/1704.01155},
  year      = {2017},
  url       = {http://arxiv.org/abs/1704.01155},
  eprinttype = {arXiv},
  eprint    = {1704.01155},
  bibsource = {dblp computer science bibliography, https://dblp.org}
}

@article{colourdepth,
title = {Color in business, science, and industry. By Deane B. Judd. John Wiley \& Sons, Inc., New York, 1952. 401 pp. Illustrated. 15.5 Ã— 24 cm. Price \$6.50},
journal = {Journal of the American Pharmaceutical Association (Scientific ed.)},
volume = {42},
number = {12},
pages = {757},
year = {1953},
issn = {0095-9553},
doi = {https://doi.org/10.1002/jps.3030421221},
url = {https://www.sciencedirect.com/science/article/pii/S0095955315330675}
}

@book{smoothing,
author = {Nixon, Mark and Aguado, Alberto S.},
title = {Feature Extraction \& Image Processing, Second Edition},
year = {2008},
isbn = {0123725380},
publisher = {Academic Press, Inc.},
address = {USA},
edition = {2nd}
}

@TECHREPORT{nonlocal,
    author = {Antoni Buades and Bartomeu Coll and Jean Michel Morel},
    title = {On image denoising methods},
    institution = {TECHNICAL NOTE, CMLA (CENTRE DE MATHEMATIQUES ET DE LEURS APPLICATIONS},
    year = {2004}
}

@article{svd,
author = {Wu, Fei and Xiao, Limin and Yang, Wenxue and Jinbin, Zhu},
year = {2020},
month = {09},
pages = {173},
title = {Defense against adversarial attacks in traffic sign images identification based on 5G},
volume = {2020},
journal = {EURASIP Journal on Wireless Communications and Networking},
doi = {10.1186/s13638-020-01775-5}
}

@Inbook{svdpca,
author="Wall, Michael E.
and Rechtsteiner, Andreas
and Rocha, Luis M.",
editor="Berrar, Daniel P.
and Dubitzky, Werner
and Granzow, Martin",
title="Singular Value Decomposition and Principal Component Analysis",
bookTitle="A Practical Approach to Microarray Data Analysis",
year="2003",
publisher="Springer US",
address="Boston, MA",
pages="91--109",
isbn="978-0-306-47815-4",
doi="10.1007/0-306-47815-3_5",
url="https://doi.org/10.1007/0-306-47815-3_5"
}

@misc{advtrain,
      title={Recent Advances in Adversarial Training for Adversarial Robustness}, 
      author={Tao Bai and Jinqi Luo and Jun Zhao and Bihan Wen and Qian Wang},
      year={2021},
      eprint={2102.01356},
      archivePrefix={arXiv},
      primaryClass={cs.LG}
}

@INPROCEEDINGS{regressionattack,
  author={Jagielski, Matthew and Oprea, Alina and Biggio, Battista and Liu, Chang and Nita-Rotaru, Cristina and Li, Bo},
  booktitle={2018 IEEE Symposium on Security and Privacy (SP)}, 
  title={Manipulating Machine Learning: Poisoning Attacks and Countermeasures for Regression Learning}, 
  year={2018},
  volume={},
  number={},
  pages={19-35},
  doi={10.1109/SP.2018.00057}}

@misc{svmattack,
      title={Poisoning Attacks against Support Vector Machines}, 
      author={Battista Biggio and Blaine Nelson and Pavel Laskov},
      year={2013},
      eprint={1206.6389},
      archivePrefix={arXiv},
      primaryClass={cs.LG}
}

@misc{moosavidezfooli2017universal,
      title={Universal adversarial perturbations}, 
      author={Seyed-Mohsen Moosavi-Dezfooli and Alhussein Fawzi and Omar Fawzi and Pascal Frossard},
      year={2017},
      eprint={1610.08401},
      archivePrefix={arXiv},
      primaryClass={cs.CV}
}

@article{REN2020346,
title = {Adversarial Attacks and Defenses in Deep Learning},
journal = {Engineering},
volume = {6},
number = {3},
pages = {346-360},
year = {2020},
issn = {2095-8099},
doi = {https://doi.org/10.1016/j.eng.2019.12.012},
url = {https://www.sciencedirect.com/science/article/pii/S209580991930503X},
author = {Kui Ren and Tianhang Zheng and Zhan Qin and Xue Liu}
}

@InProceedings{Vivek_2018_ECCV,
author = {Vivek, B. S. and Mopuri, Konda Reddy and Babu, R. Venkatesh},
title = {Gray-box Adversarial Training},
booktitle = {Proceedings of the European Conference on Computer Vision (ECCV)},
month = {September},
year = {2018}
}

@ARTICLE{8611298,
  author={Yuan, Xiaoyong and He, Pan and Zhu, Qile and Li, Xiaolin},
  journal={IEEE Transactions on Neural Networks and Learning Systems}, 
  title={Adversarial Examples: Attacks and Defenses for Deep Learning}, 
  year={2019},
  volume={30},
  number={9},
  pages={2805-2824},
  doi={10.1109/TNNLS.2018.2886017}
}

@misc{papernot2017practical,
      title={Practical Black-Box Attacks against Machine Learning}, 
      author={Nicolas Papernot and Patrick McDaniel and Ian Goodfellow and Somesh Jha and Z. Berkay Celik and Ananthram Swami},
      year={2017},
      eprint={1602.02697},
      archivePrefix={arXiv},
      primaryClass={cs.CR}
}

@inproceedings{papernot2016distillation,
  title={Distillation as a defense to adversarial perturbations against deep neural networks},
  author={Papernot, Nicolas and McDaniel, Patrick and Wu, Xi and Jha, Somesh and Swami, Ananthram},
  booktitle={2016 IEEE symposium on security and privacy (SP)},
  pages={582--597},
  year={2016},
  organization={IEEE}
}

@article{hinton2015distilling,
  title={Distilling the knowledge in a neural network},
  author={Hinton, Geoffrey and Vinyals, Oriol and Dean, Jeff},
  journal={arXiv preprint arXiv:1503.02531},
  year={2015}
}

@inproceedings{soll2019evaluating,
  title={Evaluating defensive distillation for defending text processing neural networks against adversarial examples},
  author={Soll, Marcus and Hinz, Tobias and Magg, Sven and Wermter, Stefan},
  booktitle={International Conference on Artificial Neural Networks},
  pages={685--696},
  year={2019},
  organization={Springer}
}

@article{papernot2017extending,
  title={Extending defensive distillation},
  author={Papernot, Nicolas and McDaniel, Patrick},
  journal={arXiv preprint arXiv:1705.05264},
  year={2017}
}

@misc{metzen2017detecting,
      title={On Detecting Adversarial Perturbations}, 
      author={Jan Hendrik Metzen and Tim Genewein and Volker Fischer and Bastian Bischoff},
      year={2017},
      eprint={1702.04267},
      archivePrefix={arXiv},
      primaryClass={stat.ML}
}

@misc{lu2017safetynet,
      title={SafetyNet: Detecting and Rejecting Adversarial Examples Robustly}, 
      author={Jiajun Lu and Theerasit Issaranon and David Forsyth},
      year={2017},
      eprint={1704.00103},
      archivePrefix={arXiv},
      primaryClass={cs.CV}
}

@misc{feinman2017detecting,
      title={Detecting Adversarial Samples from Artifacts}, 
      author={Reuben Feinman and Ryan R. Curtin and Saurabh Shintre and Andrew B. Gardner},
      year={2017},
      eprint={1703.00410},
      archivePrefix={arXiv},
      primaryClass={stat.ML}
}

@misc{song2018pixeldefend,
      title={PixelDefend: Leveraging Generative Models to Understand and Defend against Adversarial Examples}, 
      author={Yang Song and Taesup Kim and Sebastian Nowozin and Stefano Ermon and Nate Kushman},
      year={2018},
      eprint={1710.10766},
      archivePrefix={arXiv},
      primaryClass={cs.LG}
}

@inproceedings{NIPS2016_b1301141,
 author = {van den Oord, Aaron and Kalchbrenner, Nal and Espeholt, Lasse and kavukcuoglu, koray and Vinyals, Oriol and Graves, Alex},
 booktitle = {Advances in Neural Information Processing Systems},
 editor = {D. Lee and M. Sugiyama and U. Luxburg and I. Guyon and R. Garnett},
 pages = {},
 publisher = {Curran Associates, Inc.},
 title = {Conditional Image Generation with PixelCNN Decoders},
 url = {https://proceedings.neurips.cc/paper/2016/file/b1301141feffabac455e1f90a7de2054-Paper.pdf},
 volume = {29},
 year = {2016}
}

@misc{salimans2017pixelcnn,
      title={PixelCNN++: Improving the PixelCNN with Discretized Logistic Mixture Likelihood and Other Modifications}, 
      author={Tim Salimans and Andrej Karpathy and Xi Chen and Diederik P. Kingma},
      year={2017},
      eprint={1701.05517},
      archivePrefix={arXiv},
      primaryClass={cs.LG}
}

@Book{EfroTibs93,
  Title                    = {An Introduction to the Bootstrap},
  Author                   = {Bradley Efron and Robert J. Tibshirani},
  Publisher                = {Chapman \& Hall/CRC},
  Year                     = {1993},
  Address                  = {Boca Raton, Florida, USA},
  Number                   = {57},
  Series                   = {Monographs on Statistics and Applied Probability}
}

@misc{kurakin2017adversarial,
      title={Adversarial examples in the physical world}, 
      author={Alexey Kurakin and Ian Goodfellow and Samy Bengio},
      year={2017},
      eprint={1607.02533},
      archivePrefix={arXiv},
      primaryClass={cs.CV}
}

@misc{szegedy2014intriguing,
      title={Intriguing properties of neural networks}, 
      author={Christian Szegedy and Wojciech Zaremba and Ilya Sutskever and Joan Bruna and Dumitru Erhan and Ian Goodfellow and Rob Fergus},
      year={2014},
      eprint={1312.6199},
      archivePrefix={arXiv},
      primaryClass={cs.CV}
}

@misc{goodfellow2015explaining,
      title={Explaining and Harnessing Adversarial Examples}, 
      author={Ian J. Goodfellow and Jonathon Shlens and Christian Szegedy},
      year={2015},
      eprint={1412.6572},
      archivePrefix={arXiv},
      primaryClass={stat.ML}
}

@misc{dong2018boosting,
      title={Boosting Adversarial Attacks with Momentum}, 
      author={Yinpeng Dong and Fangzhou Liao and Tianyu Pang and Hang Su and Jun Zhu and Xiaolin Hu and Jianguo Li},
      year={2018},
      eprint={1710.06081},
      archivePrefix={arXiv},
      primaryClass={cs.LG}
}

@inproceedings{papernot2016limitations,
  title={The limitations of deep learning in adversarial settings},
  author={Papernot, Nicolas and McDaniel, Patrick and Jha, Somesh and Fredrikson, Matt and Celik, Z Berkay and Swami, Ananthram},
  booktitle={2016 IEEE European symposium on security and privacy (EuroS\&P)},
  pages={372--387},
  year={2016},
  organization={IEEE}
}

@article{simonyan2013deep,
  title={Deep inside convolutional networks: Visualising image classification models and saliency maps},
  author={Simonyan, Karen and Vedaldi, Andrea and Zisserman, Andrew},
  journal={arXiv preprint arXiv:1312.6034},
  year={2013}
}

@article{wiyatno2018maximal,
  title={Maximal jacobian-based saliency map attack},
  author={Wiyatno, Rey and Xu, Anqi},
  journal={arXiv preprint arXiv:1808.07945},
  year={2018}
}

@article{arjovsky2017towards,
  title={Towards principled methods for training generative adversarial networks},
  author={Arjovsky, Martin and Bottou, L{\'e}on},
  journal={arXiv preprint arXiv:1701.04862},
  year={2017}
}

@article{zhao2017generating,
  title={Generating natural adversarial examples},
  author={Zhao, Zhengli and Dua, Dheeru and Singh, Sameer},
  journal={arXiv preprint arXiv:1710.11342},
  year={2017}
}

@article{goodfellow2014generative,
  title={Generative adversarial nets},
  author={Goodfellow, Ian and Pouget-Abadie, Jean and Mirza, Mehdi and Xu, Bing and Warde-Farley, David and Ozair, Sherjil and Courville, Aaron and Bengio, Yoshua},
  journal={Advances in neural information processing systems},
  volume={27},
  year={2014}
}

@article{ILSVRC15,
    Author = {Olga Russakovsky and Jia Deng and Hao Su and Jonathan Krause and Sanjeev Satheesh and Sean Ma and Zhiheng Huang and Andrej Karpathy and Aditya Khosla and Michael Bernstein and Alexander C. Berg and Li Fei-Fei},
    Title = {{ImageNet Large Scale Visual Recognition Challenge}},
    Year = {2015},
    journal   = {International Journal of Computer Vision (IJCV)},
    doi = {10.1007/s11263-015-0816-y},
    volume={115},
    number={3},
    pages={211-252}
}

@misc{deng2020analysis,
      title={An Analysis of Adversarial Attacks and Defenses on Autonomous Driving Models}, 
      author={Yao Deng and Xi Zheng and Tianyi Zhang and Chen Chen and Guannan Lou and Miryung Kim},
      year={2020},
      eprint={2002.02175},
      archivePrefix={arXiv},
      primaryClass={eess.SP}
}

@misc{ramachandran2017fast,
      title={Fast Generation for Convolutional Autoregressive Models}, 
      author={Prajit Ramachandran and Tom Le Paine and Pooya Khorrami and Mohammad Babaeizadeh and Shiyu Chang and Yang Zhang and Mark A. Hasegawa-Johnson and Roy H. Campbell and Thomas S. Huang},
      year={2017},
      eprint={1704.06001},
      archivePrefix={arXiv},
      primaryClass={cs.LG}
}

@misc{meng2017magnet,
      title={MagNet: a Two-Pronged Defense against Adversarial Examples}, 
      author={Dongyu Meng and Hao Chen},
      year={2017},
      eprint={1705.09064},
      archivePrefix={arXiv},
      primaryClass={cs.CR}
}

@misc{papernot2016science,
      title={Towards the Science of Security and Privacy in Machine Learning}, 
      author={Nicolas Papernot and Patrick McDaniel and Arunesh Sinha and Michael Wellman},
      year={2016},
      eprint={1611.03814},
      archivePrefix={arXiv},
      primaryClass={cs.CR}
}

@book{10.5555/1162254, author = {Rasmussen, Carl Edward and Williams, Christopher K. I.}, title = {Gaussian Processes for Machine Learning (Adaptive Computation and Machine Learning)}, year = {2005}, isbn = {026218253X}, publisher = {The MIT Press} }

@article{janiesch2021machine,
  title={Machine learning and deep learning},
  author={Janiesch, Christian and Zschech, Patrick and Heinrich, Kai},
  journal={Electronic Markets},
  pages={1--11},
  year={2021},
  publisher={Springer}
}
\end{document}